# Low artificial anisotropy cellular automaton model and its applications to the cell-to-dendrite transition in directional solidification

Lei Wei[1, 2], Xin Lin[2*], Meng Wang[2], Weidong Huang[2]

[1]School of Mechanical Engineering, Northwestern Polytechnical University, Xi'an 710072 P.R. China

[2]State Key Laboratory of Solidification Processing, Northwestern Polytechnical University, Xi'an 710072 P.R. China

**Abstract**

A low artificial anisotropy cellular automaton (CA) model is developed for the simulation of microstructure evolution in directional solidification. The CA model's capture rule was modified by a limited neighbor solid fraction (LNSF) method. Various interface curvature calculation methods have been compared. The simulated equilibrium shapes agree with the theoretical shapes, when the interface energy anisotropy coefficient is $\varepsilon=0.01$, $\varepsilon=0.03$ and $\varepsilon=0.05$, respectively. The low artificial anisotropy CA model is used in the numerical simulation of the cell-to-dendrite transition (CDT) in directional solidification. The influence of physical parameters ($\Gamma$, $D_l$, $k_0$, $m_l$) on CDT has been investigated. The main finding in this paper is the discovery of the changing behavior of the $V_{cd}$ when the solute partition coefficient $k_0$ is larger than a critical value. When $k_0$ is less than 0.125, the $V_{cd}$ follows the Kurz and Fisher criterion $V_c/k_0$; while when $k_0>0.125$, the $V_{cd}$ equals to $8V_c$. The experimental data of succinonitrile-acetone (SCN-ace, $k_0=0.1$) and SCN-camphor ($k_0=0.33$) support the conclusion from CA simulations.

## 1. Introduction

The microsegregation between nonplanar solid-liquid interfaces strongly influences the material's mechanical properties. During the directional solidification of alloys, the solid-liquid interface can be a planar, cellular or dendritic morphology, which is depending on the growth conditions (pulling velocity $V$, thermal gradient $G$ and alloy composition $C_0$). The instability transition from a planar to a cellular interface at a low velocity and that from a cellular to a planar interface at a high velocity have been established by Mullins-Sekerka instability theory [1]. The cellular interface instability can be cell elimination, tip splitting, or side-branch emission. The side-branch emission, which is also called as the cell-to-dendrite transition (CDT), has remained poorly predicted by theories. Kurz and Fisher's theory [2] predicted that the CDT occurred at $V_{cd}=V_c/k_0$ ($V_{cd}$ = growth velocity of cellular-dendrite transition, $V_c$

---

* Corresponding author.
E-mail address: xlin@nwpu.edu.cn (X. Lin).
Phone: +86 029 88494001

= planar growth stability limit, $k_0$ = solute partition coefficient). Trivedei [3] and Somboonsuk et al. [4] have found that the CDT occurred at the minimum in the solute peclet number, which supported to the Kurz and Fisher criterion. Laxmanan [5] compared these theories with experimental data from several alloys. It was found that only the $V_{cd}$ in succinonitrile-acetone (SCN-ace) and Al-2%Cu alloys were close to $V_c/k_0$. Laxmanan suggested that the large variation of $V_{cd}$ may be due to the crystallographic anisotropy effects or the buoyancy convection. Chopra and Tewari [6] showed that the CDT appeared to be strongly influenced by the magnitude of the solute partition coefficient $k_0$. In their experiments, the $V_{cd}$ in Pb-Sn alloy ($k_0$ = 0.5) was larger than the prediction of the Kurz and Fisher criterion.

Recently, the critical spacing $\lambda_{cd}$ was introduced into the investigations of the CDT as a control parameter [8-10]. It has been found that the CDT was not sharp. The cell and dendrite coexist in a region of pulling velocities depending on the local spacing. However, the expressions of the critical spacing $\lambda_{cd}$ presented between Georgelin and Pocheau [8], and Trivedi et al. [9, 10] are quite different. Based on the expression of the $\lambda_{cd}$, Trivedi et al. also concluded that the CDT initiates when the maximum cell spacing $\lambda_{c, max}$ equals to $\lambda_{cd}$. However, the expression of $\lambda_{c, max}$ is still qualitative. Hunt and Lu [11] have presented a rough expression for cell spacing $\lambda_{c, max}$. Phase field simulations [12, 14] shown that the maximum finger spacing $\lambda_{c, max}$ was proportion to $1/\sqrt{V}$ ($V$ = the pulling velocity), which was different with the predictions by the Saffman-Taylor viscous finger problem [14]. To date, it is still a great challenge to establish the precise expressions of the $\lambda_{c, max}$, $\lambda_{d, min}$ and $\lambda_{cd}$, especially considering the influences of the interface energy coefficient $\varepsilon$.

In order to solve the problems mentioned above, numerical simulation is a good option. Numerical model has more control parameters ($V$, $G$, $C_0$, $\lambda$, $\Gamma$, $D_l$, $k_0$, $m_l$, $\varepsilon$) than that in thin film experiments ($V$, $G$, $C_0$, $\lambda$). It is able for the numerical model to quantitatively examine the influences of physical parameters. Phase field model [13, 14] has been used to simulate the CDT by controlling the primary spacing, which gave deep understandings of the CDT. Karma et al. [14] suggested to give a more exhaustive survey as a function of the various parameters ($d_0/l_T$, $l_D/l_T$, $\Lambda/l_T$, $k_0$, and $\varepsilon$), but such a survey remains a nontrivial computational challenge for phase filed model. To date, the Phase field (PF) [15] and cellular automaton (CA) [16] are the most popular computational models for the simulation of solidification microstructure. Compared to the phase field model, CA model has advantages in computational efficiency. The disadvantage of CA model is the artificial anisotropy.

The idea of CA was originally introduced by Von Neumann [17] in 1940s to reproduce complex physical phenomena with simple rules. In 1980s, Stefan Wolfram [18] discovered the classic elementary cellular automaton (Rule 110, for instance), which is capable of universal computation. The CA model has highly computational efficiency and relatively simple physical principles. Due to these advantages, CA model has been widely used in simulations of dendrite growth [19-31]. To simulate dendrite growth, Nastac [19] used Von Neumann type of neighborhood definition, the results of which shown strong artificial anisotropy. He also used a counting cell method to calculate the interface curvature. The accuracy of the counting cell method was evidenced to be dependent on mesh size [20]. So far as known, a method based on the variation of the unit vector normal (VUVN) to the solid-liquid interface along the direction of the interface is a better solution [20, 21]. However, the VUVN method needs to calculate the derivatives of solid fractions, which is difficult to be accurately calculated in a sharp interface model. The roughness in capture rule definition and curvature calculation are the origin of the artificial anisotropy in CA model.

In order to reduce the artificial anisotropy, there were mainly two kinds of CA models. One was the virtual front tracking method, presented by Zhu and Stefanescu [21]; the second was the decentred square algorithm, presented by P.D. Lee and H.B. Dong [22,23]. Both of the two methods could make the dendrites grow in arbitrary directions, which was an important advancement. However, there were empirical decision rules in the calculations of interface curvature in the virtual front tracking model. The decentred square algorithm had the disadvantage that the simulated results were still influenced by artificial anisotropy. Both of the two kinds of CA models had no quantitatively examination of the artificial anisotropy by the simulation of the equilibrium shapes. In recent years, the progressive developments of the CA model were basically based on the two kinds of the CA models. For example, the six symmetrical dendrite growth CA model could be considered as a development of the decentred square algorithm in the hexagonal grid [24]. More importantly, there were some new kinds of CA models. A.Z. Lorbiecha's PCA model [25], which was based on the randomly distributed CA Points to improve the capture rules and the curvature calculations, achieved the dendrite growth in arbitrary directions. However, it introduced a new problem that the dendritic morphology was not smooth. M. Marek [26] presented a Growth Anisotropy Reduction with Diffusion method (GARED), which could simulate a dendrite with six fold symmetry on the Cartesian square CA mesh, instead of hexagonal mesh. However, he used kinetic anisotropy instead of interface energy anisotropy. Our previous research provided a zigzag capture rule [27-29] and a limited neighbor solid fraction (LNSF) rule [30]. Both of them were designed to overcome artificial anisotropy. Overall, the modification of the capture rule in the CA model will be a long-term investigation. We also introduced a bilinear interpolation algorithm [28] to modify the derivatives of solid fractions in VUVN method. The accuracy of the interface curvature calculation was improved to a large extent.

Recently, the quantitative comparison of steady state dendrite tip velocities between the PF and CA models was presented [31]. It was recommended to use a hybrid method, which means that a CA model's outputs are as the inputs of a PF model [32]. However, it is difficult to give a comprehensive comparison between the PF and CA models. Despite the dendrite tip velocities, other solidification morphologies, such as cellular interface [33,34] should also be compared. Unfortunately, due to the artificial anisotropy, most of the CA models' outputs were dendrite morphologies. The cellular and seaweed morphologies in directional solidification require small interface energy anisotropy, which is neglected by strong artificial anisotropy in the CA model. So, the CA model has a disadvantage of the precisely describing the morphology of solidification microstructure, especially which is sensitive to interface anisotropy. To date, the PF model is the state-of-the-art numerical model for the simulation of microstructure in solidification process. The PF model implicitly captures the solid-liquid interface, based on the phase-field variable $\Phi$(solid phase for $\Phi=1$, liquid phase for $\Phi=0$, interface for $0<\Phi<1$). Using the PF variable $\Phi$, it is convenient for the PF model to accurately calculate the interface curvature. By contrast, the interface curvature is difficult to be accurately calculated in the CA model. The CA model, as a sharp interface model, uses a discontinuous Heaviside function of the solid fractions to capture the interface. The interface can be reconstructed by the straightforward SLIC method [37] or by PLIC method [38].

However, the desire of a tracking interface can also be found in the numerical simulation of multiphase computational fluid dynamics (CFD) [39-45]. There were two important approaches for the CFD to capture free interface positions: the volume-of-fluid [39] and the level-set approaches [40]. It can be seen that both of the CA method and the volume-of-fluid approach are sharp interfaces; and both of the PF method and the level-set approach are diffusive interface. In volume-of-fluid approach, the investigations to improve interface curvature have been continuously carried out for decades. Brackbill et al. [42] presented a continuum surface force (CSF) model, in which the volume-of-fluid was convolved with a smoothing kernel. Cummins et al. [43] compared the accuracy of curvature estimates derived from three volume-of-fluid based functions: a convolved volume-of-fluid function (CV), a height function (HF), and a reconstructed distance function (RDF). It was found that the curvature estimates derived from the height function provided superior results. In future work, the CA model can use the latest new curvature estimate method in volume-of-fluid approach, because both of the CA model and volume-of-fluid approach use a discontinuous Heaviside function (the volume fractions) on an Eulerian (fixed) grid to represent the interface.

In this paper, a low artificial anisotropy CA model is developed for the simulation of directional solidification. The influences of physical parameters ($\Gamma$, $D_l$, $k_0$, $m_l$ $k_0$) on the CDT in directional solidification are investigated by the present low artificial anisotropy CA model.

## 2. Numerical description of CA model

The computational domain is divided into Cartesian grid. Each grid, which is also called cell, is characterized by three states, such as liquid, solid and interface, as seen in Fig.1. In order to govern the transition of cell states, a capture rule is needed to control the evolution of different states. Solid fraction (solid cell $f_s = 1$, liquid cell $f_s = 0$, interface cell $0 < f_s < 1$) is introduced to implicitly capture the solid-liquid interface. The growth of solid fractions can be calculated according to the interface kinetics, which are based on the algorithms of the interface curvature calculation and the thermal or mass transport calculation. The thermal and mass transport calculation methods can be found elsewhere [20-23]. The following subsections focus on the capture rule, interface kinetics and interface curvature calculation.

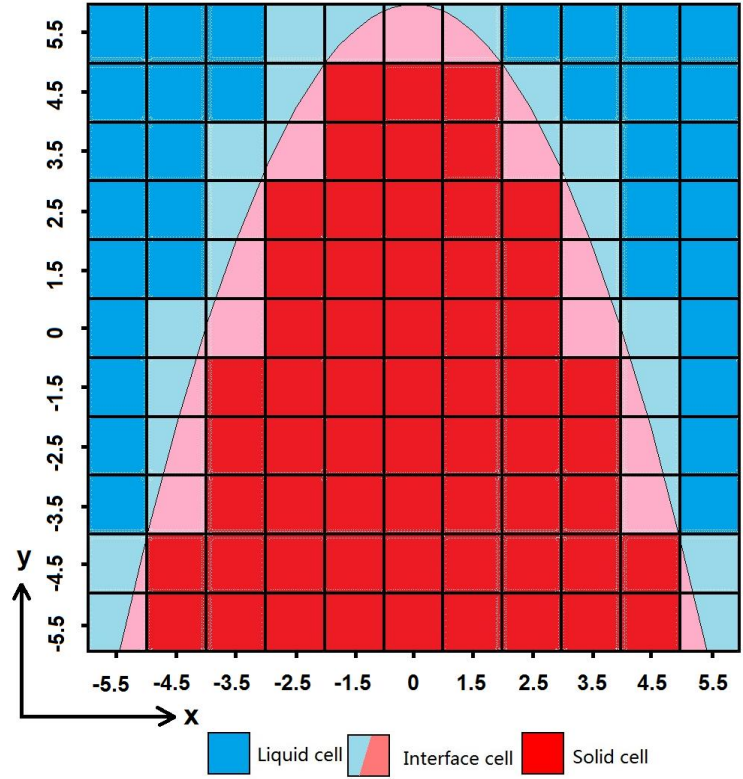

Fig. 1 The scheme of solid-liquid interface in CA model: liquid cell, interface cell and solid cell.

### 2.1 Capture rule

During the CA simulation, the transition of cell state from liquid to interface is governed by the capture rule. Since the transition of cell states influences the growth of solid-liquid interface, the capture rule used in CA model should be carefully selected. The traditional capture rules, such as Von Neumann's and Moore's rules, were evidenced to have strong artificial anisotropy [20, 28]. The capture rule in the present CA model used a limited neighbor solid fraction (LNSF) method [19], which is a modification of the Von Neumann's rule.

Based on the Von Neumann's rule,the LNSF method calculates the averaged solid fraction $fs_{ave}$ around a specific liquid cell. If $fs_{ave}$ is larger than a constant value $fs_{const}$, then the liquid cell can be captured by Von Neumann's rule, otherwise if $fs_{ave}$ is less than $fs_{const}$, the liquid cell cannot be captured even if it is satisfied by Von Neumann's rule. The LNSF method was effective for the pure substance CA mode [30], in which the artificial anisotropy could be reduced to a large extent. More importantly, the LNSF method is only based on some basic algebraic operators, which means that the LNSF method is as computational efficiency as Von Neumann's rule. In the present work, the LNSF method was applied to the alloy CA model.

### 2.2 Interface kinetics

The transition from interface cell to solid cell is determined by the interface kinetics, which governs the growth of solid fraction. At the solid-liquid interface, the temperature and concentration should satisfy the following expressions:

$$T^* = T_l^{eq} + (C_l^* - C_0)m_l - \Gamma K f(\varphi, \theta_0) \tag{1}$$

$$C_s^* = k_0 C_l^* \tag{2}$$

where, $T^*$ is the temperature at the interface, $T_l^{eq}$ is the melting point at the initial composition $C_0$, $C_l^*$ and $C_s^*$ are the interface compositions in solid and liquid phases, respectively, $m_l$ is the liquidus slope, $\Gamma$ is the Gibbs–Thomson coefficient, $K$ is the interface curvature, $f(\varphi,\theta_0)=1-15\varepsilon\cos(4(\varphi-\theta_0))$, in which $\varepsilon$ is the interface energy anisotropy coefficient, $\varphi$ is the growth angle between the normal to the interface and the x-axis, $\theta_0$ is the angle of the preferential growth direction with respect to the x-axis. In the 3D CA model, interface curvature is calculated by Hoffman-Cahn [30, 46, 47] $\xi$-vector.

The interface growth kinetics used in the present work are proposed by Zhu and Stefanescu [21]:

$$C_l^* = C_0 - \frac{T_l^{eq} - T^* - \Gamma K f(\varphi, \theta_0)}{m_l} \tag{3}$$

$$\Delta f_s = (C_l^* - C_l^{old}) / (C_l^*(1 - k_0)) \tag{4}$$

$$C_l^{new} = C_l^{old} / (1 - (1 - k_0) * \Delta f_s) \tag{5}$$

where, $C_l^{new}$ and $C_l^{old}$ are the actual concentrations at different time steps.

## 2.3 Interface curvature calculation

The lack of accuracy in curvature calculation has significant influence on the accuracy of the CA model. From the local equilibrium condition, Eq. (1), we can see that if the calculation of interface curvature $K$ is inaccurate, the curvature undercooling can't reflect the changes with interface energy anisotropy. So far as known, the most popular method for the simulation of interface curvature is the counting cell method [19], Eq. (6). However, the counting cells method is not accurate enough for quantitative simulation [20].

$$K = \frac{1}{\Delta x}\left(1 - 2\frac{fs + \sum_{i=1}^{n} fs(i)}{n+1}\right) \tag{6}$$

A more accurate method is based on the variation of the unit vector normal (VUVN) to the solid-liquid interface along the direction of the interface, Eq. (7) [20, 21].

$$K = \frac{2\frac{\partial fs}{\partial x}\frac{\partial fs}{\partial y}\frac{\partial^2 fs}{\partial x \partial y} - \left(\frac{\partial fs}{\partial y}\right)^2 \frac{\partial^2 fs}{\partial x^2} - \left(\frac{\partial fs}{\partial x}\right)^2 \frac{\partial^2 fs}{\partial y^2}}{\left[\left(\frac{\partial fs}{\partial x}\right)^2 + \left(\frac{\partial fs}{\partial y}\right)^2\right]^{3/2}} \tag{7}$$

The VUVN method needs to calculate the derivatives of solid fraction. They are difficult to be precisely calculated in a sharp interface model. Both of the CA and volume-of-fluid algorithms were introduced certain types of interpolation methods to accurately calculate the derivatives of solid fraction (volume fraction in the volume-of-fluid method). The interpolation method used in the CA model was based on bilinear interpolation, the detail of which can be found in reference [27, 28].

One of the interpolation method used in volume-of-fluid algorithm is described as follows [39]. The subscript $\{x, i+1/2, j+1/2\}$ denote the partial derivative with respect to x at $\{i+1/2, j+1/2\}$:

$$fs_{x,i+1/2,j+1/2} = \frac{1}{2dx}(fs_{i+1,j} - fs_{i,j} + fs_{i+1,j+1} - fs_{i,j+1}) \tag{8}$$

$$fs_{y,i+1/2,j+1/2} = \frac{1}{2dy}(fs_{i,j+1} - fs_{i,j} + fs_{i+1,j+1} - fs_{i+1,j}) \tag{9}$$

$$fs_{x,i,j} = \frac{1}{4}(fs_{x,i+1/2,j+1/2} + fs_{x,i-1/2,j+1/2} + fs_{x,i+1/2,j-1/2} + fs_{x,i-1/2,j-1/2}) \tag{10}$$

$$fs_{y,i,j} = \frac{1}{4}(fs_{y,i+1/2,j+1/2} + fs_{y,i-1/2,j+1/2} + fs_{y,i+1/2,j-1/2} + fs_{y,i-1/2,j-1/2}) \tag{11}$$

We used the solid fractions divided by a parabolic interface in Fig. 1 to test the four curvature calculation methods: the counting cell method, the VUVN method without interpolation, the VUVN method based on bilinear interpolation and the VUVN method in volume-of-fluid approach. Fig. 2 is the comparison results of the four curvature calculation methods. From the results, we can see that the methods of counting cell and the VUVN method without interpolation algorithm were not as accurate as the other two methods. The VUVN method in volume-of-fluid approach was slightly better than the VUVN method with bilinear interpolation. We used the former method in the following simulations of dendrite growth. In order to gain more accurate calculation results of interface curvature, new methods should be introduced to the CA model in future investigations, such as a convolved volume-of-fluid function (CV), a height function (HF) or a reconstructed distance function (RDF) [41-45].

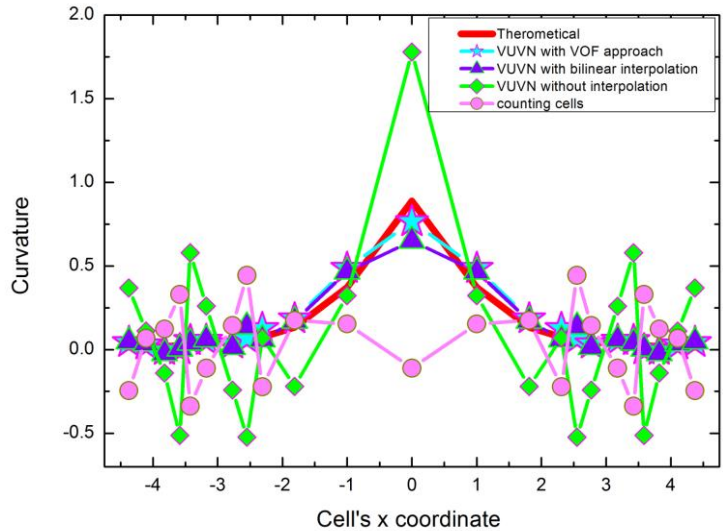

Fig. 2 The comparison of various interface curvature calculation methods to the theoretical results.

## 3   Results and discussion

All the simulations in the present paper were run on a personal computer with processors of AMD Phenom 3.30GHz without parallel simulation. 2D simulations of directional solidification were within 2 hours each. 3D simulations of directional solidification were less than 24 hours each.

### 3.1   Elimination of the artificial anisotropy in the present CA model

The artificial anisotropy in the CA model was qualitative examined in the previously developed pure substance models [27-30]. According to Karma [35], the quantitative capability of the PF model was examined by the growth of equilibrium shapes. In the present work, we also used equilibrium shapes to quantitatively verify the present CA model's artificial anisotropy. It was the first time for the CA model to simulate equilibrium shapes.

The alloy used in the simulations is SCN-0.4mol%acetone. The isothermal solidification model was used here. The boundary conditions of the concentration field were set with zero flux conditions. The thermophysical properties of SCN-acetone can be found elsewhere [4]. The simulation of an equilibrium shape was started with a solid circle seed (seed radius equals to 23 μm) in bulk melt with small undercooling (less than 0.003K). After certain number of time steps (more than 400000 steps, 6 seconds of real time, mesh size of 1μm), the circular interface slowly grew into four fold symmetry due to the interface energy anisotropy coefficient $\varepsilon$. Fig.3 is the comparison of equilibrium shapes under different interfacial energy anisotropy coefficient: $\varepsilon=0.01$, $\varepsilon=0.03$ and $\varepsilon=0.05$, respectively. The theoretical plot is according to the Cahn-Hoffman ξ-vector for a model fourfold anisotropy given by $1+\varepsilon*\cos(4\theta)$ [36]. Fig.3(a) is the results by present CA model, in which the LNSF capture rule and a VUVN with VOF approach curvature calculation method are used. Fig.3(b) is the results of traditional CA model, in which the Von Neumann capture rule and a counting cell curvature calculation method are used. It can be seen that the equilibrium shapes simulated by the present CA model were agreed well with the theoretical equilibrium shapes. The equilibrium shapes simulated by the traditional CA model are not agree with the theoretical equilibrium shapes. Actually, the changing of the interface energy anisotropy coefficient $\varepsilon$ from 0.01-0.05 has no effects on the simulated morphologies by traditional CA model. The reason is that the counting cell method cannot calculate accurate interface curvature. Inaccurate interface curvature will neglect the influence of the interface energy anisotropy coefficient $\varepsilon$. The results in Fig.3 have shown that our modifications to the CA model (LNSF capture rule, curvature calculation based on interpolation algorithm) had very positive effects on the elimination of the CA model's artificial anisotropy.

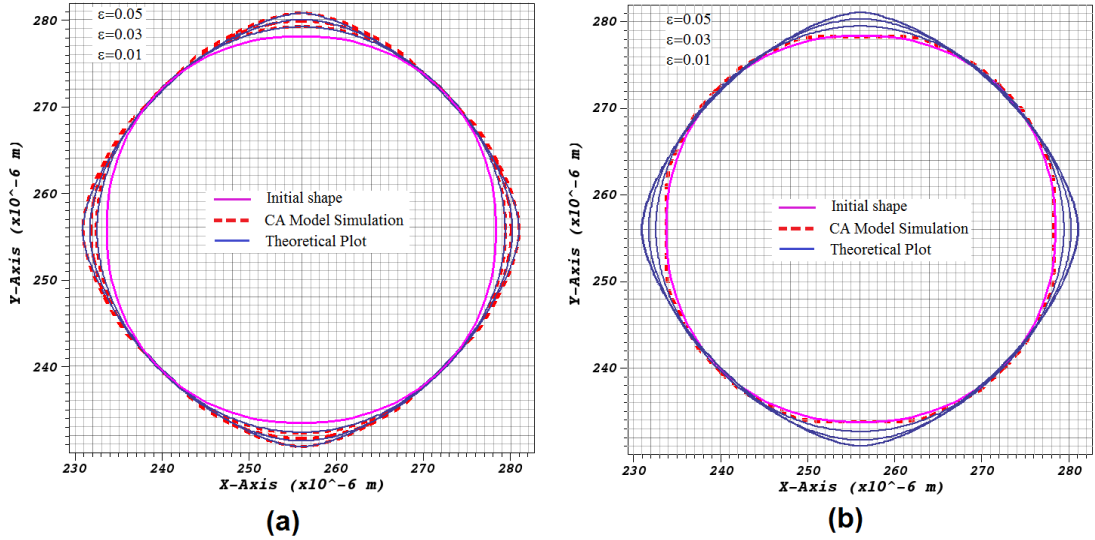

Fig. 3 The simulated equilibrium shapes under different interface energy anisotropy coefficients, and the comparison to the theoretical equilibrium shapes:(a) the LNSF capture rule with VUVN-VOF curvature calculation method; (b) the Von Neumann capture rule with counting cell curvature calculation method.

**3.2 The critical primary spacing of the cell-to-dendrite transition in directional solidification**

Due to the artificial anisotropy, CA simulations of cellular growth in directional solidification were found very few in literatures. Since the present CA model has low artificial anisotropy, the cellular and CDT interface morphologies in directional solidification were presented, which have never been simulated by CA model.

The alloy used in the simulation of directional solidification is succinonitrile-0.1mol%acetone (SCN-0.1mol%ace). The computational domain is 384μm×1536 μm for 2D model, and the mesh size is 1.5 μm. If the domain is filled with regular grid with mesh size of 1.5 μm, the domain can be divided into 256×1024 grids. At the beginning of the simulations, the bottom of the computational domain is initialed as a planar interface. Fixed temperature gradient is pulled up along the longer side of the domain.

Under constant pulling velocity $V$ =100 μm/s, the growth morphologies under different temperature gradients are shown in Fig.4. The temperature gradients $G$ from Fig.4 (a)-(e) are 0.5 K/mm, 5 K/mm, 10 K/mm, 15 K/mm and 20 K/mm, respectively. The simulated morphologies were dendrites when $G$<15 K/mm. It was cellular when $G$=20 K/mm. In between, when $G$=15 K/mm, the simulated morphology was the CDT.

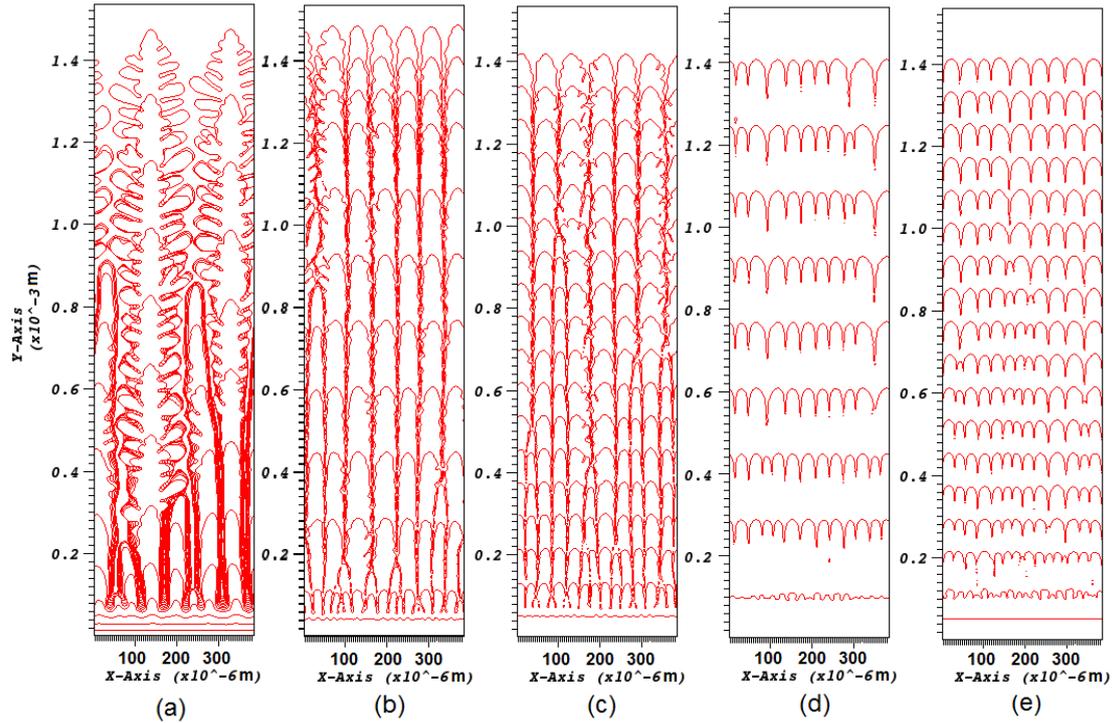

Fig. 4 Time evolution of the morphology from a planar interface, the interface energy anisotropy $\varepsilon =0.005$, the pulling velocity $V$=100 μm/s, and the temperature gradients: (a) $G$=0.5 K/mm; (b) $G$= 5 K/mm, (c) $G$=10 K/mm, (d) $G$=15 K/mm, (e) $G$=20 K/mm.

We analyzed the changes of various values at the cell/dendrite tips from planar interface instability to steady state cell/dendrite arrays under different temperature gradients, as seen in Fig.5. Fig.5 (a) is the tip velocities versus time. All the tip velocities converged to 100 μm/s, except the tip velocity at $G$=0.5 K/mm, which was still needed more time to be steady state. Fig.5 (b) is the changes of tip concentrations. As the increasing of temperature gradient $G$ from 0.5 K/mm to 20 K/mm, the steady state tip concentration also increased. Fig.5 (c) is the tip temperature changing with time. The tip temperature was decreased as the increasing of temperature gradient.

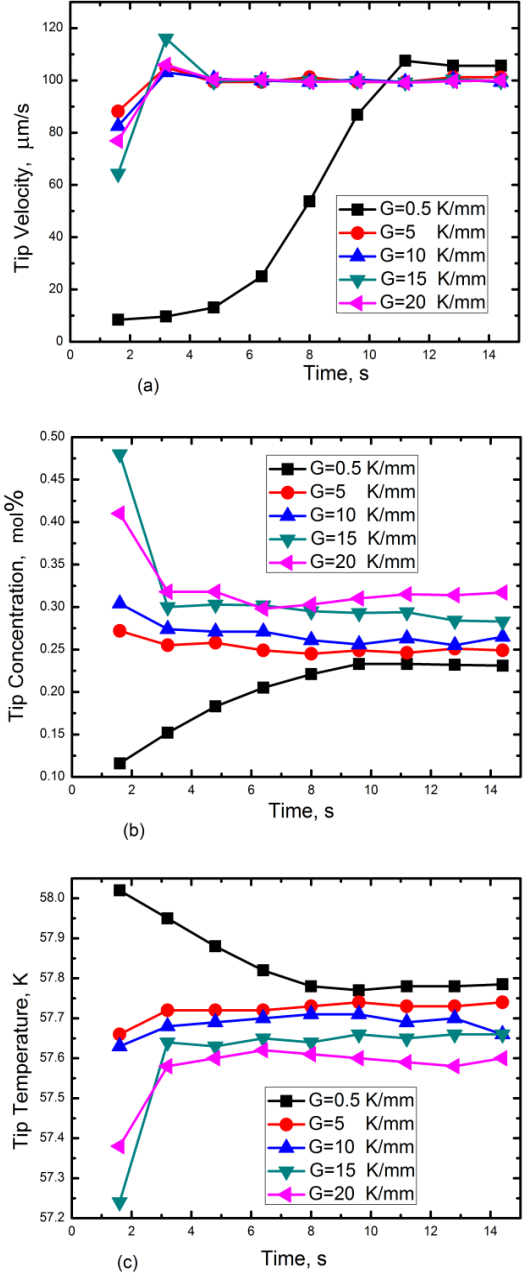

Fig. 5 The changes of quantities at the cell/dendrite tips during planar interface to steady state cell/dendrite arrays: (a) tip velocities; (b) tip concentrations; (c) tip temperatures.

According to the Kurz and Fisher criterion [2,3], the CDT occurs under the conditions described by Eq. (12)

$$l_D = k_0 l_T \tag{12}$$

where, $l_D$ is solute diffusion length, $l_T = \Delta T_0/G$ is thermal diffusion length.

Tthe Kurz and Fisher criterion is experimentally evidenced to give well predictions for the CDT of SCN-ace alloy[3-5]. When $V$=100 μm/s, the temperature gradient for the CDT to occur calculated by Eq. (12) is $G$=15.256 K/mm, which is very close to the temperature gradient in Fig.4 (d). The CDT in the present CA model was agreed with the theory expectation. In order to quantitatively investigate the CDT, critical primary spacing $\lambda_{cd}$ should be carefully calculated [8-10]. We focused on the Fig.4 (d), as seen in Fig.6. It could be obviously distinct the cellular and dendritic morphology. The primary spacing $\lambda_c$ of cell is less than the primary spacing $\lambda_d$ of dendrite. The relationship of the primary spacing can be described as $\lambda_d > \lambda_{cd} > \lambda_c$. Fig.6 (a) is the concentration map of Fig.4 (d) at the end of the simulation. Fig.6 (b) is the plot of concentrations of the three lines as demonstrated in Fig.6 (a). It can be seen that the cellular tip concentration (dotted green line) is larger than the dendritic tip concentration (dashed purple line). The dendritic tip is in front of the cellular tip, which means that the dendritic tip temperature is larger than the cellular tip temperature. In Fig.6 (a), there is an eliminated cell between two dendrites. The sidewise instability occurred when the local spacing became larger than critical value $\lambda_{cd}$.

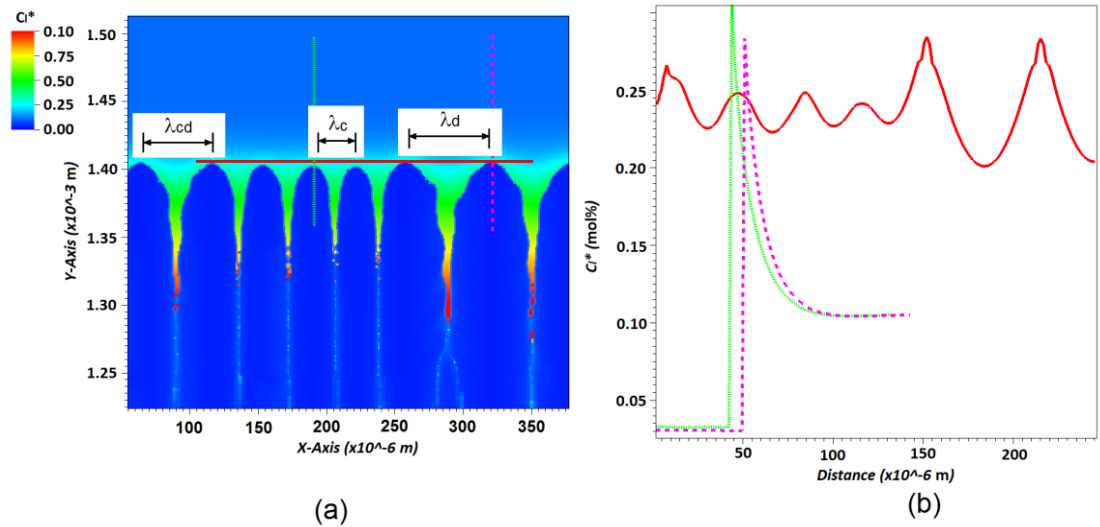

Fig. 6 Cell and dendrite morphologies of the CDT: (a) solute map; (b) plots of the concentration lines in (a)

J.Teng, S. Liu and R. Trivedi [10] presented an expression for the critical spacing for side-branch initiation in dilute SCN-acetone alloys by the investigation of the CDT through thin film experiment:

$$\lambda_{cd} = 7.63(D\Gamma)^{1/3}(GV)^{-1/3}C_0^{-1/4} \qquad (13)$$

where, $C_0$ is measured by wt.%, SCN-0.1mol%acetone equals SCN-0.0725 wt.%acetone.

Based on the simulation conditions, $\lambda_{cd}$ = 56.3 μm is obtained by Eq. (13), and $\lambda_{cd}$ = 54.5 μm is the simulated result by the present CA model in Fig.6. The present CA model has a good agreement with Eq. (13).

In order to comprehensively compare the Eq. (13) and the present CA model, we simulated the CDT by two alloys, SCN-0.05mol%acetone and SCN-0.1mol%acetone, under the temperature gradients of 5 K/mm, 10 K/mm, 15 K/mm and 20 K/mm, respectively. The corresponding pulling velocities during the simulations were slightly larger than the velocities calculated by Eq. (12). The comparison between simulation and Eq. (13) can be seen in Fig.7. It can be seen that the simulation results were agreed well with Eq. (13). It is worth noting that the $GV$ in J.Teng's experiment were between (0.001, 0.1) K/s. And the $GV$ in the CA simulations were between (0.1-6.0) K/s. For relatively large $GV$, the CA model is still agreed with Eq. (13).

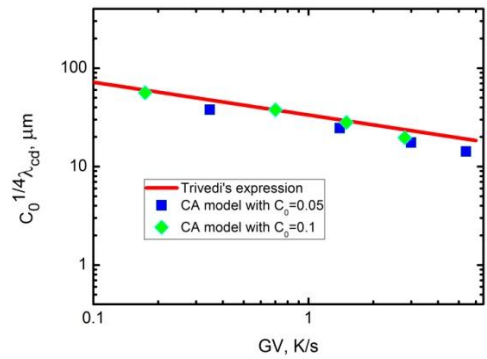

Fig. 7 The critical primary spacing of CDT, comparison between CA simulation and Eq.(13): $\lambda_{cd}C_0^{1/4}$ vs. $GV$

Besides the 2D CA model, we also developed a 3D CA model to investigate the CDT in directional solidification. The computational domain is 192μm×192μm×768μm. The mesh size is the same as the 2D CA model. Fig.8 is the simulation results of SCN-0.1mol%acetone at $\varepsilon = 0.005, V$=100 μm/s, $G$=0.5 K/mm and $G$=20 K/mm, respectively. Fig.8 (a) has the same simulation conditions as Fig.4 (a), and Fig.8 (b) is corresponding to Fig.4 (e). Fig.9 is the top view of 3D simulations, which have the same simulation conditions as that in Fig.4. The transition from dendritic to cellular microstructure could be clearly seen in the 3D directional solidification.

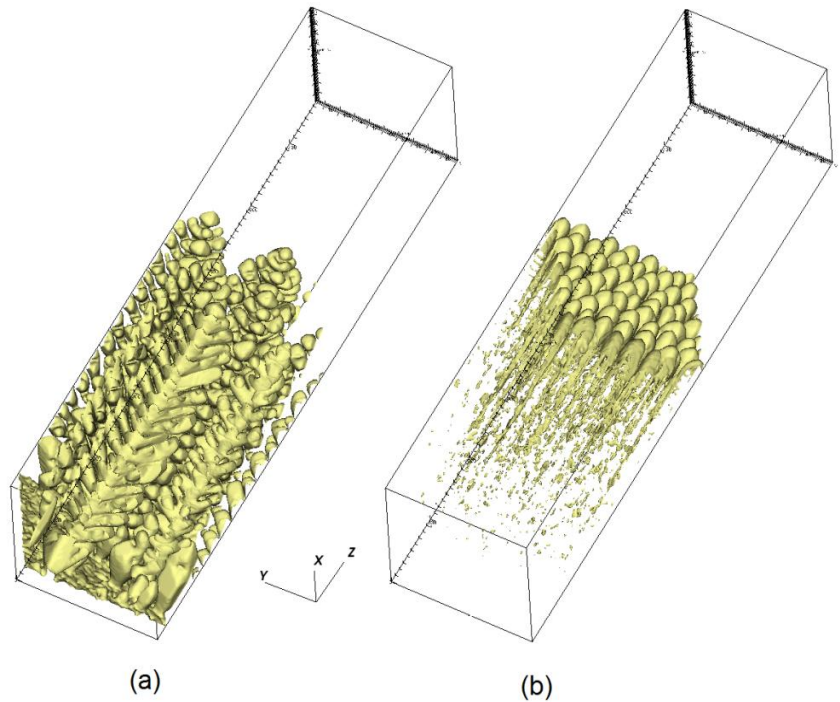

Fig. 8 Three dimensional CA model simulations of directional solidification: (a) dendrite microstructure at G=0.5 K/mm; (b) cellular microstructure at G=20 K/mm.

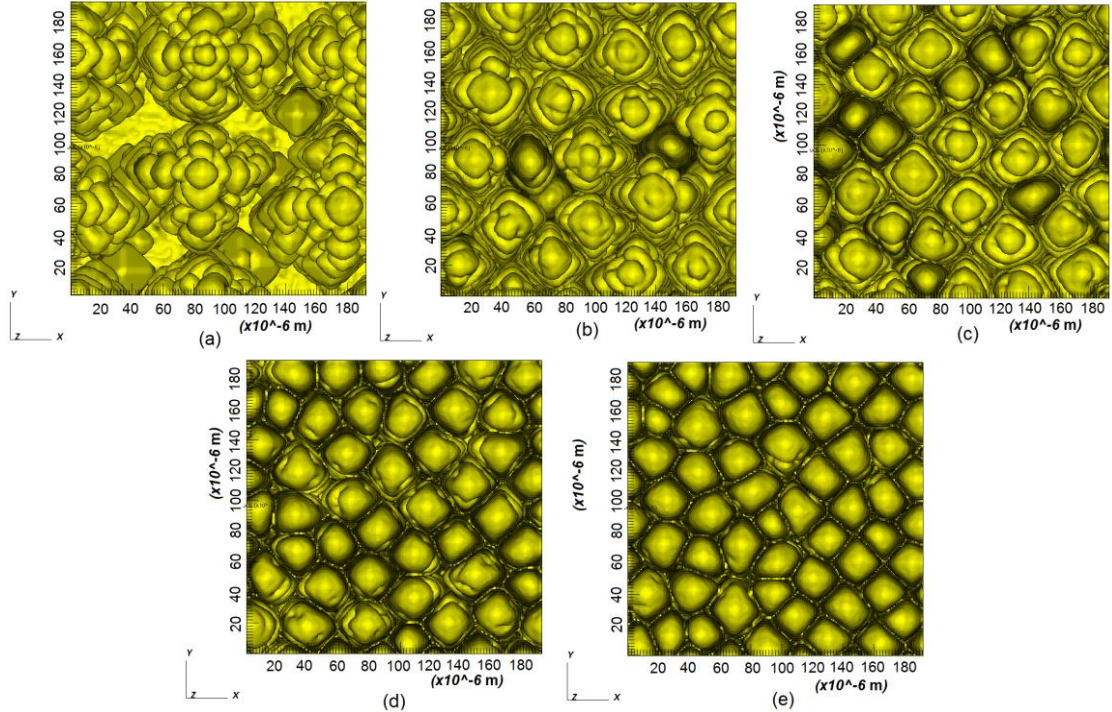

Fig. 9 The top view of the three dimensional CDT in directional solidification, the temperature gradient are: (a) G=0. 5 K/mm; (b) G= 5 K/mm, (c) G=10 K/mm, (d) G=15 K/mm, (e) G=20 K/mm, respectively.

The origin of sidebranches is caused by noise amplification or a limit cycle, which is still a standing issue in solidification. Echebarria and Karma [13] have found that the CDT cannot be understood by the phase field model without taking into account thermodynamical noise. However, many CA models[21-24], including the present low artificial anisotropy CA model, the dendritic sidebranches can be initialed without the introduction of thermodynamical noise. The dendritic sidebranches in present CA model, as seen in Fig.6 (a), is discontinuous increase in primary spacing. The origin of dendritic sidebranches in present CA model is not so-called "tail-instability". The "tail-instability" generate new cells from the sidebranches of dendrite arrays when increasing the pulling velocity[50]. Fig.4 shows that as the temperature gradient decreases (the same effect as increasing pulling velocity), the simulated morphologies continuously transit from cells to dendrites. During this process, the "tail-instability" is not observed. In the present article, we are more concerned with the control parameters of the sidebranches. The control parameters, including primary spacing $\lambda_{cd}$, temperature gradient $G$ and pulling velocity $V$ are agreed well with experiment and theory. In the next section, the influences of physical parameters on the CDT are investigated, which cannot be investigated by experiments.

## 3.3 The influences of physical parameters on the cell-to-dendrite transition in directional solidification

The alloy used in the simulations is also succinonitrile-0.1mol%acetone (SCN-0.1mol%ace). The computational domain is 256μm×2048μm for 2D CA model. If the simulation needs larger domain, the computational domain can be enlarged into 512μm×4096μm. The mesh size is 1.0 μm, in order to get more quantitative results. Instead of using fixed cell/dendrite spacing [13, 14], the cell/dendrite spacing in the present simulations is selected by the growth conditions ($V$, $G$, $C_0$), which is the same as the experiment in directional solidification.

Fig.10 (a) shows the simulation results with different strengths of Gibbs-Thomson coefficient $\Gamma$, under unchanged conditions of $V$=100 μm/s, $G$=15 K/mm, $C_0$=0.1 mol%, $m_l$=-2.16 K/mol%, $k_0$=0.103 and ε=0.005. When $\Gamma$ increases from $3.2 \times 10^{-8}$ to $9.6 \times 10^{-8}$ m∗K, both of the cell and dendrite primary spacings increase. It can be obviously seen that the changing of $\Gamma$ have no effects on the CDT. The Kurz and Fisher criterion also predict that $\Gamma$ has no effects on the CDT.

By using the same strategy, the liquidus slope $m_l$, the initial composition $C_0$ and the solute diffusivity $D_l$ are also examined by the CA simulations, as seen from Fig.10 (b) to Fig.10 (d). The corresponding changes in pulling velocities $V$ or temperature gradient $G$ are calculated according to the Kurz and Fisher criterion. The simulation results show that the three physical parameters: $m_l$, $C_0$ and $D_l$ linearly affect pulling velocities $V$ or temperature gradient $G$ of the CDT, which is also agree well with the prediction of the Kurz and Fisher criterion.

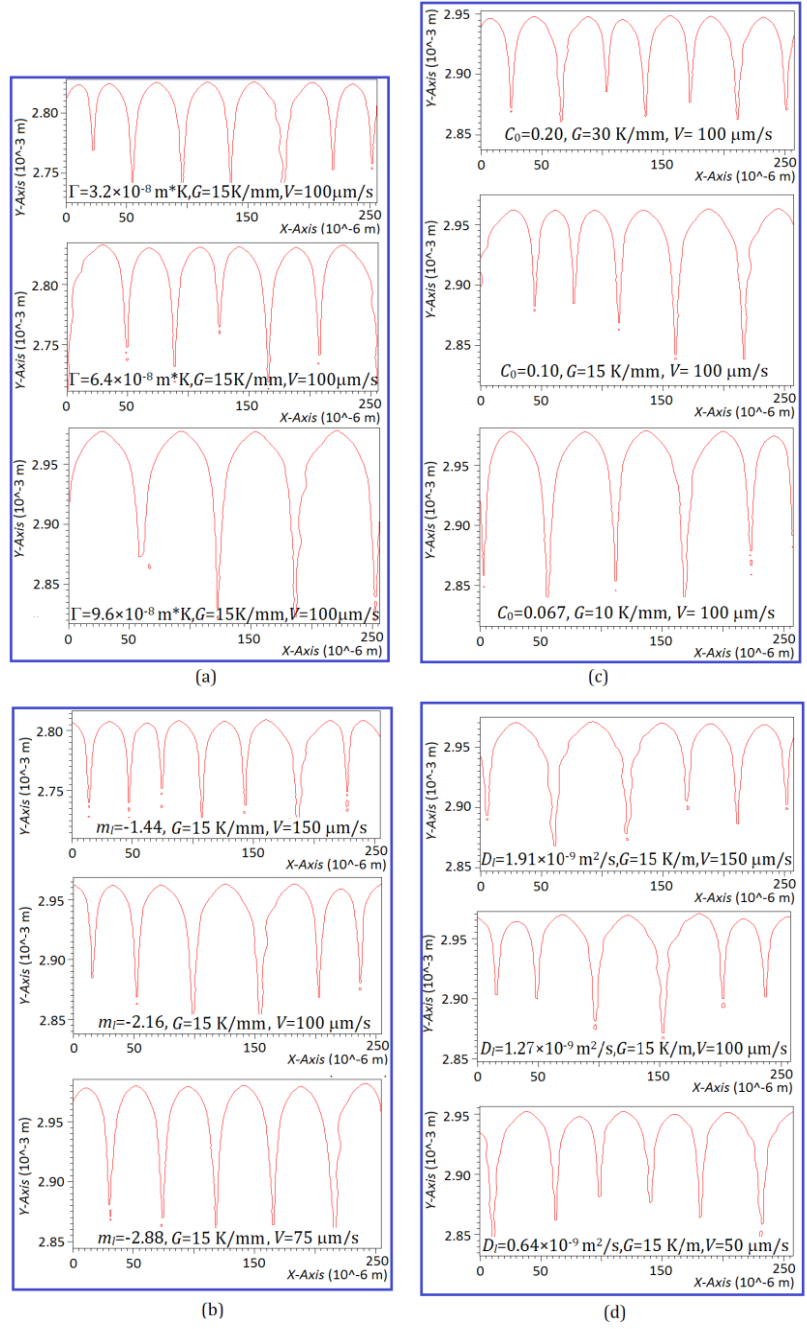

Fig. 10 The simulated CDT morphologies with different strengths of control parameters: (a) Gibbs-Thomson coefficient; (b) the liquidus slope $m_l$, while keeping $m_l*V$ constant; (c) the initial composition $C_0$, while keeping $G/C_0$ constant. (d) the solute diffusivity $D_l$, while keeping $D_l/V$ constant.

The main finding in this paper is the changing behavior of the $V_{cd}$ when the solute partition coefficient $k_0$ is larger than a critical value of 0.125. The $V_{cd}$ is defined as the smallest growth velocity at which the side branches were observed, which is the same as the experimental definition [6]. When $k_0$ is less than 0.125, the $V_{cd}$ follows the Kurz and Fisher criterion $V_c/k_0$; when $k_0>0.125$, the $V_{cd}$ equals to $8V_c$, as seen in Fig.11. It can be concluded that the occurrence of the CDT is determined by the larger value between $V_c/k_0$ and $8V_c$.

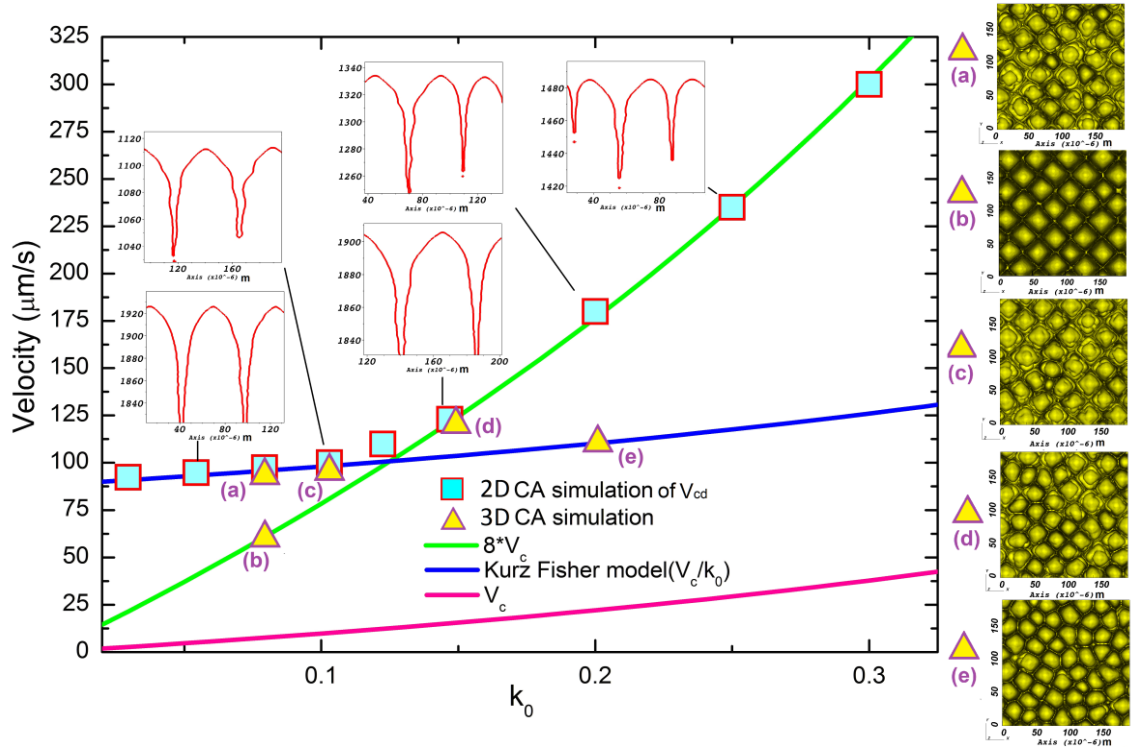

Fig. 11 The smallest velocity at which the side branches ($V_{cd}$) as a function of $k_0$. Green line: $8V_c$. Blue line: $V_c/k_0$. Magenta line: $V_c$. Squares: 2D simulation results. Triangles (a) to (e): 3D simulation results.

For the proof of this statement, 3D simulations are required, because phase field results have shown that there was a major difference between two- and three-dimensional configurations [13]. Since 3D simulations need much more computational resources than 2D simulations, only a few points have been simulated by 3D CA model. The 3D simulations show that cellular morphologies were obtained at $8V_c$ ($k_0=0.07$ in Fig.11 (b)) and $V_c/k_0$ ($k_0=0.20$ in Fig.11 (e)). Fig.11 (a), (c) and (d) are the CDT morphologies. The 3D simulation results agree with the predictions in 2D simulations.

The tip undercoolings of the cells and dendrites simulated by the 2D CA model are shown in Fig.12. The green sphere/triangle dots are the cellular/dendritic tip undercoolings as a function of $k_0$, with fixed pulling velocities $V_p=V_c/k_0$. The CDT occurred when $k_0$ was less than 0.125. The magenta sphere/triangle dots are the cellular/dendritic tip undercoolings with fixed pulling velocities $V_p=8V_c$. The CDT occurred when $k_0$ was larger than 0.125. The dendrite tip undercoolings are always smaller than the cellular tip undercoolings. By the comparison between the cellular tip undercoolings (green and magenta sphere dots), it can be seen that there is a crossover at $k_0=0.125$. The CDT (green and magenta triangle dots) occurred at which the cellular tip undercooling was smaller. It can also be concluded that the CDT behavior has a major change when $k_0=0.125$.

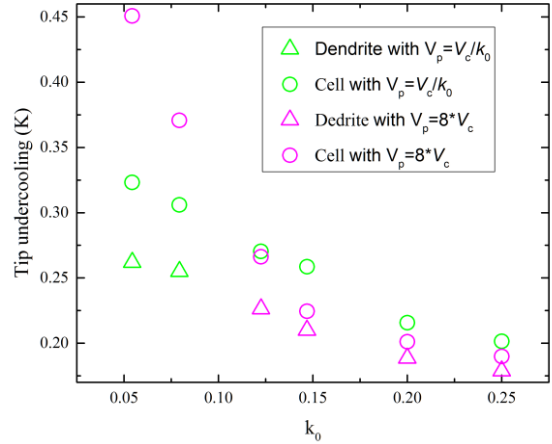

Fig. 12 Tip undercooling as a function of $k_0$. Green open spheres: cell tip undercoolings at pulling velocity $V_p=V_c/k_0$. Green open triangles: dendrite tip undercoolings at $V_p=V_c/k_0$. Magenta open spheres: cell tip undercoolings at $V_p=8V_c$. Magenta open triangles: dendrite tip undercoolings at $V_p=8V_c$.

From the experimental point of view [3-6], only the CDT behaviors in SCN-ace ($k_0$=0.1) and Al-Cu ($k_0$=0.14) alloys agreed with the Kurz and Fisher criterion. Chopra and Tewari [6] considered that the CDT appeared to be strongly influenced by the magnitude of $k_0$, but the reason for this behavior was not understood. The $V_{cd}$ in Pb-Sn alloy from Chopra and Tewari's experiment was about $2.5V_c$, which is larger than the $V_{cd}$ predicted by the Kurz and Fisher criterion, smaller than $8V_c$ from the present CA simulation results. The density differences between Pb and Sn are large, in which the convection effects would raise some doubts about the validity in Pb-Sn alloy's results [3].

The influence pattern of $k_0$ can be validated by the comparison of experimental results between SCN-ace ($k_0$=0.1) and SCN-camphor ($k_0$=0.33). SCN-ace and SCN-camphor have similar strength of the interface energy anisotropy, which makes the comparison more focused on the $k_0$. It was also evidenced in our research that the influences of the physical parameters ($D_l$, $m_l$) on the $V_{cd}$ exactly followed the Kurz and Fisher criterion, as seen in Fig.10. The differences of $m_l$ and $D_l$ in SCN-ace and SCN-camphor alloys have no effects on the comparison between $V_{cd}$ expressions influenced by $k_0$. Furthermore, both of the experiment data in SCN-camphor and SCN-ace were obtained by the same directional solidification apparatus, which is similar to that described by Hunt [48].

Trivedi et al. [10] have presented detailed experimental data of the CDT in SCN-camphor, which was charactered by $C_0$, $G$ and $V$. The smallest velocity at which the side branches ($V_{cd}$) are shown in the reference's Table 1. The experiment data in the reference's Table 1, when $C_0$=0.65wt.% and $C_0$=0.90wt.% are neglected here, because the $V_{cd}$ was not linearly changed with $G$. Another reason is that for fixed $G$, the $V_{cd}$ at $C_0$=0.65 is less than that at $C_0$=0.90. According to Kurz and Fisher criterion and our previous CA simulation, when $G$ is fixed, the $V_{cd}$ is proportional to $1/C_0$.

The $V_{cd}$ versus temperature gradient $G$ when $C_0$=0.35 is shown in Fig.13. With fixed $C_0$=0.35, the $V_{cd}$ was linearly changed with temperature gradient $G$. The experiment data are agreed with $8V_c$ better than $V_c/k_0$. Despite the neglected data, the experimental data in SCN-camphor support well to the conclusion derived from present CA simulations. For SCN-ace alloy, it was experimentally evidenced that the $V_{cd}$ follows $V_c/k_0$ [3-5]. Overall, the experiment data of SCN-camphor and SCN-ace alloys support the influence pattern of $k_0$ on the CDT discovered in this paper.

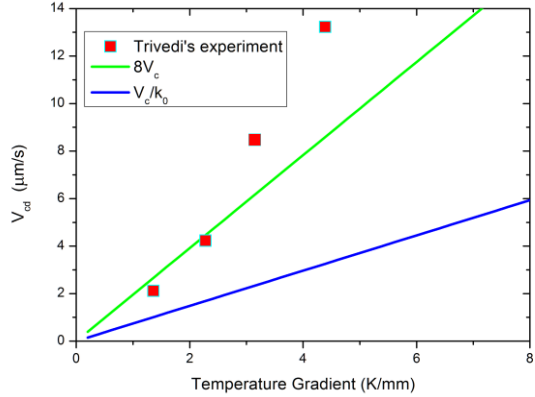

Fig. 13 The smallest velocity at which the side branches ($V_{cd}$) as a function of temperature gradient. Green line: $8V_c$. Blue line: $V_c/k_0$. Squares: Trivedi's experimental results.

Why the influence pattern of $k_0$ on the CDT has not been discovered before? Although Chopra and Tewari [6] noticed that the CDT appeared to be strongly influenced by the magnitude of the solute partition coefficient $k_0$, it cannot be sure that the difference in $V_{cd}$ is caused by $k_0$. Other thermal properties between two alloys are also different, especially the interface energy anisotropy coefficient. Phase field simulation costs much more computational recourses than CA model. Lan [12] found that the simulation from deep cells to dendrites (side branching) still remained a great challenge. Karma *et* al. [13, 14] considered that a survey of physical parameters on the CDT remained a nontrivial computational challenge for phase filed model. The development of low artificial anisotropy CA model [27-30] brings another numerical model to simulate the CDT.

## 4  Conclusions

In this paper, we have given an alloy CA model describing the microstructure in directional solidification. The CA model's capture rule was modified by a limited neighbor solid fraction (LNSF) method. Various interface curvature calculation methods have been compared. The results have shown that the variation of the unit vector normal (VUVN) method with interpolation algorithm is more accurate. We have presented the simulation results of equilibrium shapes for the testing of the artificial anisotropy in the present CA model. The simulated equilibrium shapes were at good agreements with theoretical shapes, when the interface energy anisotropy coefficient was $\varepsilon$=0.01, $\varepsilon$=0.03 and $\varepsilon$=0.05, respectively.

The cell-dendrite transition (CDT) during directional solidification has been well investigated by the present CA model. Our simulated results of the CDT in directional solidification support the expression of critical spacing ($\lambda_{cd}$), which was analyzed in the SCN-acetone system from experiment results. Comparing the results in 2D and 3D simulations, it was found that the CDT occurred at the same conditions in 2D and 3D directional solidification.

It is evidenced that the solute partition coefficient $k_0$ strongly influence the critical velocity $V_{cd}$ of the CDT. 2D CA model shown that when $k_0$ is less than 0.125, the $V_{cd}$ follows the Kurz and Fisher criterion $V_c/k_0$; while when $k_0>0.125$, the $V_{cd}$ equals to $8V_c$. 3D CA simulation and carefully selected experimental results of SCN-camphor alloy [10] support the discovery mentioned above. The discovery of the influence pattern of $k_0$ on the CDT explains why only the experimental data of the $V_{cd}$ in SCN-ace ($k_0=0.1$) and Al-Cu ($k_0=0.14$) alloys agreed with the Kurz and Fisher criterion. Other alloys, such as Pb-Sn ($k_0=0.5$) and SCN-camphor ($k_0=0.33$), have larger $V_{cd}$ than the Kurz and Fisher criterion.

The physical background of the influence pattern of $k_0$ on the CDT is still unknown. However, the weakly non-linear stability analysis of planar interface [49] shown that a subcritical bifurcation occurs when $k_0<0.45$, whereas a supercritical bifurcation is predicted when $k_0>0.45$. However, the critical value of $k_0$ on the CDT equals to 0.125, which is not the same as that in the weakly non-linear stability analysis of planar interface instabilities. It is worthy to note that Kurz and Fisher made a simplification that $k_0 \approx 0$, while deriving the criterion of $V_c/k_0$. Consequently, the Kurz and Fisher criterion fits the $V_{cd}$ well at the region $k_0$ close to 0. However, it is a theoretical challenge to derive an expression with $k_0 \neq 0$.

**Acknowledgements**

This work was supported by the National Natural Science Foundation of China (Grant Nos. 51271213), the National Basic Research Program ("973" Program) of China (No. 2011CB610402) and the Specialized Research Fund for the Doctoral Program of Higher Education of China (Grant No. 20116102110016). This work was also supported by the China Postdoctoral Science Foundation (2013M540771).

**References**

[1] W.W. Mullins, R.F. Sekerka, J. Appl. Phys, 35, 444 (1964)

[2] W. Kurz, D.J. Fisher, Acta Metall, 29, 11 (1981).

[3] R Trivedi, Metall. Trans. A, 15A, 977 (1984).

[4] K. Somboonsuk, J. T. Mason, and R. Trivedi: Metall. Trans. A, 15A, 967 (1984).

[5] S.N. Tewari, V. Laxmanan, Metall. Trans. A, 18, 167 (1987).

[6] M.A. Chopra, S.N. Tewari, Metall. Trans. A, 22, 2467 (1991).

[7] A. Karma, P. Pelce, Europhysics Letters, 9, 713 (1989).

[8] M. Georgelin, A. Pocheau, Phys. Rev. E 57, 3189 (1998).

[9] R. Trivedi, Y. Shen, S. Liu, Metall. Mater. Trans. A 34, 395 (2003).

[10] J. Teng, S. Liu, R. Trivedi, Acta. Mater. 57, 3497 (2009)

[11] J.D. Hunt, S.Z. Lu, Metal Mater Trans A, 27, 611 (1996).

[12] C.W. Lan, C.J. Shih, M.H. Lee. Acta. Mater. 53, 2285 (2005).

[13] B. Echebarria, A. Karma, S. Gurevich, Physical Review E, 81, 021608 (2010).

[14] S. Gurevich, A. Karma, M. Plapp, R. Trivedi, Physical Review E, 81, 011603 (2010).

[15] Boettinger, Warren, Beckermann, Karma, Annual Review of Materials Research 32, 163 (2002)

[16] K. Reuther, M. Rettenmayr, Com. Mater. Sci. 95, 213 (2014)

[17] J. Von Neumann, A.W. Burks. Theory of self-reproducing automata. Urbana, University of Illinois Press (1966).

[18]     S. Wolfram, Nature, 311, 419 (1984).

[19]     L. Nastac, Acta. Mater. 47, 4253 (1999).

[20]     L.B. Sanchez, D.M. Stefanescu, Metall. Mater. Trans. A 34, 367 (2003).

[21]     M.F. Zhu, D.M. Stefanescu, Acta. Mater. 55, 1741 (2007).

[22]     W. Wang, P.D. Lee, M. McLean. Acta. Mater. 51, 2971 (2003).

[23]     H.B. Dong, P.D. Lee, Acta. Mater. 53, 659 (2004).

[24]     H. Yin, S.D. Felicelli, Modelling Simul. Mater. Sci .Eng. 17, 075011 (2009).

[25]     A.Z. Lorbiecka, B. Šarler, IOP Conf. Ser.: Mater. Sci. Eng. 27, 012057 (2012).

[26]     M. Marek, Physica D 253, 73 (2013).

[27]     X. Lin, L. Wei, M. Wang, W.D. Huang, Mater. Sci. Forum PRICM 7, 1528 (2010).

[28]     L. Wei, X. Lin, M. Wang, W.D. Huang, App. Phys. A 103, 123 (2011).

[29]     L. Wei, X. Lin, M. Wang, W.D. Huang, Com. Mater. Sci. 54, 66 (2012).

[30]     L. Wei, X. Lin, M. Wang, W.D. Huang, Physica B 407, 2471 (2012).

[31]     A. Choudhury, K. Reuther, E. Wesner, A. August, B. Nestler, M. Rettenmayr, Com. Mater. Sci., 55, 263 (2012).

[32]     W.D. Tan, N.S. Bailey, Y.C. Shin, Com. Mater. Sci., 50, 2573 (2011).

[33]     M. Plapp, J. Cryst. Growth 303 49 (2007).

[34]     C.W. Lan, C.J. Shih, M.H. Lee. Acta. Mater. 53, 2285 (2005).

[35]     A. Karma, W.J. Rappel, Phys. Rev. E 57, 4324 (1998).

[36]     A. A. WHEELER, Proc. R. Soc. A 462, 3363 (2006)

[37]     W. F. Noh and P. Woodward, in Proceedings, Fifth International Conference on Fluid Dynamics, edited by A. I. Vande, Vooren and P. J. Zandbergen, Lecture Notes in Physics,Vol. 59 (Springer-Verlag, Berlin, 1976), p. 330.

[38]     D. L. Youngs, Time dependent multimaterial flow with large fluid distortion, in Numerical Methods for Fluid Dynamics, edited by K. M. Morton and M. J. Baines (Academic Press, New York, 1982), p. 27.

[39]     D. Gueyffier, J. Li, A. Nadim, R. Scardovelli, S. Zaleski, Journal of Computational Physics, 152, 423 (1999).

[40]     M. Sussman, E.G. Puckett, Journal of Computational Physics, 162, 301 (2000).

[41]     S. Afkhami, M. Bussmann, Int. J. Numer. Meth. Fluids, 57, 453 (2008).

[42]     J.U. Brackbill, D.B. Kothe, C. Zemach, Journal of Computational Physics, 100, 335 (1992).

[43]     S.J. Cummins, M.M. Francois, D.B. Kothe, Computers & Structures, 83, 425 (2005).

[44]     D. Gerlach, G. Tomar, G. Biswas, F. Durst, Int J Heat Mass Trans, 49, 740 (2006).

[45]     H.S. Udaykumar, L. Mao, Int J Heat Mass Trans, 45, 4793 (2002).

[46]     D.W. Hoffman, J.W. Cahn, Surf. Sci. 31, 368 (1972).

[47]     J.E. Taylor, Acta. Metall. Mater. 40, 1475 (1992).

[48]     J.D. Hunt, K. A. Jackson, and H. Brown: Rev. Sci. Instrum., 37, 805 (1966).

[49]     B. Caroli, C. Coroli, and B. Roulet:J. Phys., 43, 1767 (1982).

[50]     Y. Saito, C. Misbah, H. Muller-Krumbhaar, Phys. Rev. L. 63, 2377 (1989)